\documentclass[12pt]{article}
\usepackage{amsfonts}
\usepackage{bbm}
\usepackage{graphicx}
\usepackage{booktabs}
\usepackage{subfigure}
\usepackage{mathrsfs}
\usepackage{color}
\usepackage{verbatim}
\usepackage{cancel}
\usepackage{geometry}             
\usepackage{amssymb}
\usepackage{amsmath}
\usepackage{epstopdf}
\usepackage{cases}
\usepackage{cite}
\usepackage{setspace}

\usepackage[hyperindex=true,
pdfstartview=FitH,
bookmarksnumbered=true,
bookmarksopen=true,
citecolor=blue,
linkcolor=blue,
colorlinks=true,
urlcolor=rossoCP3,
unicode]{hyperref}

\definecolor{rossoCP3}{cmyk}{0,.88,.77,.40}

\begin{document}

\title{\bf
Revisit on thermodynamics of BTZ black hole with variable Newton constant}
{\author{\small Yan-Ying Bai${}^{1}$, Xuan-Rui Chen${}^{1}$, Zhen-Ming Xu${}^{1,2,3}$, Bin Wu${}^{1,2,3}$\thanks{{\em email}: \href{mailto:binwu@nwu.edu.cn}{binwu@nwu.edu.cn}}{ }
        \vspace{5pt}\\
        \small $^{1}${\it School of Physics, Northwest University, Xi'an 710127, China}\\
        \small $^{2}${\it Shaanxi Key Laboratory for Theoretical Physics Frontiers, Xi'an 710127, China}\\
        \small $^{3}${\it Peng Huanwu Center for Fundamental Theory, Xi'an 710127, China}
    }
	
\date{}
\maketitle
\begin{spacing}{1.15}
\begin{abstract}
The thermodynamics of the BTZ black holes are revisited with variable Newton constant. A new pair of conjugated variables, the central charge $C$ and the chemical potential $\mu$, is introduced as thermodynamic variables.  The first law of thermodynamics and the Euler relation, instead of the Smarr relation in the extended phase space formalism, are matched perfectly in this formalism. Compatible with standard extensive thermodynamics, the black hole mass is verified to be a first order homogeneous function of the related extensive variables, and restores the role of internal energy. In addition, the heat capacity has also resulted in a first order homogeneous function in this formalism as we expected, and an asymptotic behavior in high temperature limit is shown intriguingly. The non-negatively of heat capacity indicates that the rotating and charged BTZ black holes are thermodynamically stable.
\end{abstract}

\section{Introduction}
Half a century ago, the black hole, one of the most fantastic predictions in general relativity, was suggested that it should be regarded as a thermodynamic system in the pioneering works of Hawking and Bekenstein \cite{Bekenstein:1973ur,Hawking:1976de}. Bekenstein argued that the entropy of the black hole should be proportional to its area of horizon, and Hawking calculated the thermal spectrum of the black hole with a Hawking temperature in semi-classical approach. The analogy between the ordinary thermodynamic system and the black hole thermodynamics was subsequently analyzed, the four laws of the black hole thermodynamics were established\cite{Bardeen:1973gs}. In recent decades, particular attention have been paid to the thermodynamics of the AdS black hole because of the special role of the AdS spacetime in the AdS/CFT correspondence \cite{Maldacena:1997re}.

To interpret the non-trivial contribution from the non-vanishing cosmological constant in the case of AdS black hole, the extended phase space black hole thermodynamics was developed\cite{Kastor:2009wy,Dolan:2010ha,Dolan:2011xt,Kubiznak:2012wp,Cai:2013qga}, which brings black hole thermodynamics to a new stage, and now it is also called black hole chemistry \cite{Kubiznak:2016qmn}. The central idea of the extended phase space thermodynamics is the introduction of a new thermodynamic pair pressure-volume of the black hole, where the negative cosmological constant is treated as the thermodynamic pressure of the black hole, and its conjugate variable is the thermodynamic volume of the black hole. Many efforts have been done in the study of the black hole thermodynamics in extended phase space by using the standard thermodynamic formalism, and abundant thermodynamics behaviors, such as phase transition and critical behaviors for the AdS black hole are discovered, see \cite{Zhao:2013oza,Wei:2012ui,Altamirano:2013ane,Xu:2014kwa,Frassino:2014pha,Dehghani:2014caa,Hendi:2017fxp,Wang:2018xdz,Xu:2015rfa} for examples and references therein. Besides, it should be noticed that the mass of the black hole is interpreted as enthalpy rather than the internal energy because of the existence of $VdP$ term in the first thermodynamic law in the context of the extended phase space.

The black hole thermodynamics in extended phase space attracts lots of interests, and then a question naturally arises in the framework of holography, i.e., what are the means of variation of the cosmological constant in the boundary field theory? According to the AdS/CFT correspondence, it is argued that the change of cosmological constant in gravity theory means the change of the number of colors $N$ or degree of freedom $N^2$ (which is related to the central charge $C$) in boundary theory\cite{Kastor:2014dra,Caceres:2015vsa,Zhang:2014uoa}. Alternatively, it is also suggested that if one needs to keep the boundary field theory fixed in a certain situation, one could vary the cosmological constant $\Lambda$ as well as vary Newton constant $G$, provided that the ratio $C\sim l^{d-2}/G$ keeps as a constant \cite{Karch:2015rpa}, where $l$ is AdS radius. In recent Visser's work, holographic thermodynamics was proposed, in which the central charge is introduced as a novel thermodynamics variable, and the Euler relation, as well as the first law of thermodynamics of the boundary field theory are obtained \cite{Visser:2021eqk}. Furthermore, there are also some interesting attempts to consider the contributions of holographic central charges to the black hole thermodynamics in extended phase space \cite{Cong:2021jgb,Cong:2021fnf}.

Despite the rapid developments in black hole thermodynamics since the proposal of the extended phase space, there are still some unaddressed issues that need further investigation. For instance, the thermodynamic volume conjugating to the thermodynamic pressure of the black hole lacks physical interpretation, although sometimes its value coincidentally equal to the geometric volume inside the horizon of the black hole. Moreover, the well-known Smarr relation of the rotating, charged AdS black holes for the $d$-dimensional extended phase space thermodynamics is \cite{Frassino:2015oca}
 \begin{align}
 (d-3)M&=(d-2)T S-2 P V+(d-3)\Phi Q + (d-2) \Omega J   \label{5},
 \end{align}
which shows us that the thermodynamic potential $M$ can not be written as a homogeneous function  for its arguments $S, P, Q, J$ with universal order. However, it is very important for the presence of the Euler homogeneity in standard thermodynamics, which plays a crucial role in understanding the equilibrium thermodynamic states, including those of the black holes. To address the questions mentioned above, a restricted version of Visser's holographic thermodynamics is proposed in Gao and Zhao's work\cite{Zeyuan:2021uol}, in which the cosmological constant is considered to be fixed and the Newton constant still varies, consequently, the first law of thermodynamics and the Euler homogeneity are matched perfectly in this restricted phase space thermodynamics formalism. This formalism has been verified further in the more general cases and higher dimensional spacetimes\cite{Gao:2021xtt,Zhao:2022dgc,Kong:2022gwu}. 

In three-dimensional spacetime, it was once thought that there was no black hole solution until the well-known BTZ black hole was discovered in the Einstein-gravity theory with negative cosmological constant\cite{Banados:1992wn,Martinez:1999qi}. This class of three dimensional black holes attracted much attention due to their special applications in the AdS$_3$/CFT$_2$. The thermodynamics of the BTZ black holes have been extensively studied in the extended phase space formalism\cite{Akbar:2011qw,Hendi:2016hbe,Gunasekaran:2012dq,Liang:2017kht,Frassino:2019fgr}. Nevertheless, recalling the Smarr relation Eq.(\ref{5}), the lack of the Euler homogeneity becomes more prominent, i.e., the left hand side of the Eq.(\ref{5}) is identically equal to zero and also the term $\Phi Q$ vanishes regardless of the existence of the electric-magnetic field. This ``reduced'' Smarr relation of the three dimensional black hole confuses us, and largely motivates us to consider the restricted thermodynamics formalism for the BTZ black holes. It would be significant to show that the three dimensional black holes also satisfy the requirements of the standard thermodynamics, where the first law of thermodynamics and Euler relation are consistent completely.

In the next section, we study the thermodynamics of rotating and charged BTZ black holes in the restricted phase space explicitly. The thermodynamic quantities are calculated and proved that they are divided into two groups: extensive and intensive in this formalism. Moreover, the thermodynamic stability of the BTZ black hole is analyzed from the behaviors of the heat capacity. Finally, a conclusion is made in Section \ref{III}. In this paper, the unit we adopted is $\hbar=c=k_B=1$, and $\varepsilon=1/\mu_0=2\pi$ to match the Gauss's law in three dimensional spacetime.

\section{Extensive thermodynamics of the BTZ black hole}\label{II}
The action of the three dimensional Einstein-gravity with Maxwell electromagnetic field is
$$
I=\int d^{3} x \sqrt{-g}\left(\frac{R-2 \Lambda}{2\kappa}-\frac{1}{4\mu_{0}} F_{\mu \nu} F^{\mu \nu}\right),\label{04}
$$
in which $\kappa=8\pi G$ is the gravity coupling constant, $\Lambda$ is the cosmological constant, $F_{\mu \nu}$ is the electromagnetic field tensor defined as $F_{\mu \nu}= \nabla_{\mu}A_{\nu}-\nabla_{\nu}A_{\mu}$.
The coupled field equations are derived by variation of the action, which gives
$$
G_{\mu \nu}-\Lambda g_{\mu \nu}=\kappa T_{\mu \nu},~~~\nabla_{\nu}F^{\mu \nu}=0,
$$
where $T_{\mu \nu}$ is the energy-momentum tensor of the electromagnetic field
$$
T_{\mu \nu}=\frac{1}{\mu_{0}}\left(F_{\mu \rho} F_{\nu \sigma} g^{\rho \sigma}-\frac{1}{4} g_{\mu \nu} F^{2}\right).
$$
 The ansatz for the line element of a rotating spacetime is\cite{Cadoni:2007ck}
\begin{align*}
	d s^{2}=-f(r) d t^{2}+f^{-1} d r^{2}+r^{2}\left(d \theta-\frac{4 G J}{r^{2}} d t\right)^{2},
\end{align*}
and the Maxwell field $A_{\mu}$ is written as
\begin{align*}
A_{\mu}=(-\Phi(r),0,0).
\end{align*}
The metric function and the electric potential can be directly obtained from the field equation
\begin{align}
f(r)=-8 GM+\frac{r^2}{l^2}+\frac{2 G Q^2\mu_{0}}{\pi}\ln\left(\frac{r}{l}\right)+\frac{16 G^{2} J^{2}}{r^{2}},~~~\Phi(r)=-\frac{\mu_{0}Q}{2\pi}\ln\left(\frac{r}{l}\right),\label{14}
\end{align} 
where $M$ is the black hole mass, $Q$ is the charge and $J$ is the angular momentum. To avoid confusion from the comparison between the metric function Eq.($\ref{14}$) and those in other references, the coupling constant $\mu_{0}$ is left here for the moment. In the following sections, the permeability $\mu_0$ is valued as $2\pi$ for simplification.

We see that the electrostatic potential of charged BTZ black holes asymptotically diverges with a logarithmic term. Usually, there are two schedules to deal with the divergence: $(1)$ A new thermodynamic
parameter associated with the renormalization length scale is introduced \cite{Frassino:2015oca}; $(2)$ A renormalized black hole mass is brought \cite{Cadoni:2007ck}, which is the approach we used in this paper.

\subsection{The rotating BTZ black hole}
For the case of rotating BTZ black hole with $J\neq0,\,Q=0$, the metric function reduces to the form
\begin{align*}
f(r)=-8 G M+\frac{r^{2}}{l^{2}}+\frac{16 G^{2} J^{2}}{r^{2}},
\end{align*}
which is very similar to that of the four and higher dimensional rotating AdS black holes. This implies the possibility to realize Euler homogeneity and the first law of thermodynamics simultaneously of the rotating BTZ black hole. 

The mass $M$ is determined by $f({r_{0}})=0$ with the radius of the event horizon $r_{0}$, it yields
\begin{align}
M=\frac{r_{0}^{2}}{8 G l^{2}}+\frac{2GJ^{2}}{r_{0}^{2}}. \label{6}        
\end{align}
The entropy, angular velocity and temperature 
of the rotating BTZ black holes in three dimensional spacetime are obtained as follows,
\begin{align*}
S&=\frac{\mathcal{A}}{4G}=\frac{\pi r_{0}}{2G},\\
\Omega&=\left(\frac{\partial M}{\partial J}\right)_{S,l}=\frac{4GJ}{r_{0}^{2}}, \\
T&=\frac{f'(r_{0})}{4\pi}=\frac{r_{0}}{2 \pi l^{2}}-\frac{8G^{2}J^{2}}{\pi r_{0}^{3}},
\end{align*}
where $\mathcal{A}=2\pi r_0$ is the horizon area.

In addition, there are another essential thermodynamics quantities for us to realize the Euler homogeneity of the black hole, which is denoted as chemical potential $\mu$ and its conjugate central charge $C$. The definition of conjugate central charge comes from the concept of the AdS/CFT correspondence, and for the three dimensional Einstein gravity it reads $C=l/8G$. Indeed we know that there are several candidates for a generalized central charge in arbitrary odd ($d-1$)-dimensional conformal field theory, they are both scaling as $l^{d-2}/G$ with an ambiguous coefficient\cite{Myers:2010tj}. The value of central charge from bulk and boundary can be normalized to match the holographic dictionary\cite{Hung:2011nu}. However, as we can see that the coefficient doesn't matter since the pair $\{\mu, C\}$ appears in the first law, and therefore only the scaling $l/G$ is important.

The central charge $C$ is treated as a novel thermal quantity in black hole thermodynamics, and we will notice that it plays the role of the amount of substance as in the standard thermodynamic system. What's more, there is a correspondence between the partition function of CFT and gravity theory in asymptotically AdS spacetime $Z_{CFT}=Z_{AdS}$ \cite{Gubser:1998bc,Witten:1998qj}, thus we promptly arrive at
\begin{align*}
\mu C=F=-T\ln Z_{CFT} = -T \ln Z_{AdS}=TI_E,
\end{align*}
where we have used the definition that the thermal partition function of the CFT at finite temperature is associated with the free energy, and the gravity partition function is calculated by the on-shell Euclidean action $I_E$. So that the chemical potential $\mu$ can be directly obtained from the on-shell Euclidean action $I_E$ of gravity, which gives by\cite{Eune:2013qs}
\begin{align*}
&I_{E H}=-\frac{1}{2 \kappa} \int d^{3} x \sqrt{g}\left(R-2\Lambda\right), \\
&I_{G HY}=-\frac{1}{\kappa} \int_{\partial M} d^{2} x \sqrt{h} K, \\
&I_{c t}=\frac{1}{\kappa} \int_{\partial M} d^{2} x \sqrt{h} (\frac{1}{l}),
\end{align*}
where $I_{EH}$ is Euclidean version of the Einstein-Hilbert action, $I_{GHY}$ is the Gibbons-Hawking-York action, $h$ is the reduced metric of the hypersurface, $K$ is the trace of the extrinsic curvature, and $I_{ct}$ is the counterterm to cancel the divergence. The result is
\begin{align*}
I_{E}=I_{EH}+I_{GHY}+I_{ct}=\beta (-\frac{r_{0}^2}{8Gl^2}+\frac{2GJ^2}{r_{0}^2}),
\end{align*}
where $\beta=1/T$ is the inverse temperature. Combining all the results of the thermodynamic quantities, it is easy to verify
\begin{align*}
TI_{E}&=\mu C =M-TS-\Omega J.\label{12}
\end{align*}
Instead of obtaining the Smarr relation of black hole thermodynamics in the extended phase space, we derive an inspiring relation which is similar to the Euler relation of the standard thermodynamic system 
\begin{align}
M=TS+\Omega J+\mu C.
\end{align}
Furthermore, the first law of thermodynamics for rotating BTZ black holes can be verified straightforwardly
\begin{align}
 \mathrm{d} M &=T \mathrm{d} S+\Omega \mathrm{d} J+\mu \mathrm{d} C\label{13}. 
\end{align}
It should be pointed out that in the proof of the first law Eq.(\ref{13}), the AdS radius $l$ is restricted as a constant and the Newton’ s constant $G$ is allowed to vary, so that this scheme is also called as the black hole thermodynamics in restricted phase space.

Next, to reveal the features of the restricted formalism of the black hole thermodynamics obviously, we would like to show that all the thermal quantities of the BTZ black hole are classified into two groups, one is extensive, and another is intensive as those in a standard extensive thermodynamic system. There exist a group of independent thermodynamic agreements $S,J,C$, and all the thermal potentials can be written as homogeneous function clearly of these independent agreements. At first, we rewrite $G,r_{0}$ as
 \begin{align}
 G=\frac{l}{8C},  ~~~ r_{0}=\frac{Sl}{4\pi C}\label{E20}.
 \end{align}
 Inserting relation Eq.(\ref{E20}) into Eq.(\ref{6}), the mass $M$ is rewritten as 
 \begin{align}
 &M=\frac{S^2}{16\pi^2 lC}+\frac{4\pi^2J^2C}{lS^2}, \label{mass}
 \end{align}	
 and the conjugated thermodynamic variables $T,\Omega,\mu$ are expressed as
 \begin{align}	 
 &T=\left(\frac{\partial M}{\partial S}\right)_{J, C}=\frac{S}{8 \pi^{2} l C}-\frac{8 \pi^{2} J^{2} C}{l S^{3}},\label{E10} \\
 &\Omega=\left(\frac{\partial M}{\partial J}\right)_{S, C}=\frac{8 \pi^{2} J C}{l S^{2}},\label{E11} \\
 &\mu=\left(\frac{\partial M}{\partial C}\right)_{S, J}=-\frac{S^{2}}{16 \pi^{2} l C^{2}}+\frac{4 \pi^{2} J^{2}}{l S^{2}}.\label{E12}
 \end{align}

 In mathematics, an $m$-th order homogeneous function with respect to its agreements $(x_{1},\dots,x_{n})$ satisfies
\begin{align*}
    f(\lambda x_{1},\dots,\lambda x_{n})=\lambda^m f(x_{1},\dots,x_{n}),  \qquad
    \sum_{i=1}^{n} x_{i} \frac{\partial f}{\partial x_{i}}=m f,
\end{align*}
which requires $m=1$ for the extensive quantities and $m=0$ for the intensive quantities in the standard thermodynamics system. From Eq.(\ref{mass}) and Eqs.(\ref{E10}-\ref{E12}), it is clear that $M$ is first order homogeneous function and $T,\Omega,\mu$ are zeroth order homogeneous function of $S,J,C$, which are the intriguing results we anticipated.
 
 To be more concrete, we would like to show that $S$ and $J$ are extensive variables. From Eq.(\ref{E11}), we have
 \begin{align}
 J=\frac{lS^2\Omega}{8\pi^2C}.\label{E13}
 \end{align} 
 Substituting Eq.(\ref{E13}) into Eq.(\ref{E10}), we get 
 \begin{align}
 T=\frac{S-l^2S\Omega^2}{8l\pi^2C}.\label{E14}
 \end{align}
 Because $S\ge0$ and $T\ge0$, we get a boundary
 \begin{align}
 \Omega \le\frac{1}{l},
 \end{align}
which helps us to solve the entropy directly from Eq.(\ref{E14})
 \begin{align}
 S=\mathcal{S}C,~~~~\mathcal{S}=\frac{8\pi^2lT}{1-l^2\Omega^2}.\label{E16}
 \end{align}
 We see that $S$ is proportional to $C$ and the coefficient $\mathcal{S}$ only depends on the intensive variables $T$ and $\Omega$, proving that $S$ is an extensive variable.
 
 Inserting Eq.(\ref{E16}) into Eq.(\ref{E13}), $J$ is also shown to be an extensive variable
 \begin{align*}
 J=\mathcal{J}C,~~~~\mathcal{J}=\frac{8\pi^2l^3T^2\Omega}{(1-l^2\Omega^2)^2},
 \end{align*} 
 where the coefficient $\mathcal{J}$ only depends on the intensive variables $T$ and $\Omega$.
 
 Moreover, the Gibbs-Duhem equation can be written down from Eq.(\ref{13}) and Eq.(\ref{12})
 \begin{align*}
 \mathrm{d} \mu=-\mathcal{S} \mathrm{d} T-\mathcal{J} \mathrm{d} \Omega,
 \end{align*}
 where $\mathcal{S}=S / C, \mathcal{J}=J / C$ are zeroth order homogeneous functions of $S, J, C$. The Gibbs-Duhem equation suggests that $\mu$ is dependent on $\Omega$ and $T$, indicating $\mu$ is not an independent variable. Combining Eqs.(\ref{E10}-\ref{E12}), it can be shown the partial derivative $\mu$ with respect to $T$ at constant $\Omega$ is straightforward to get   
\begin{equation}
        \left(\frac{\partial \mu}{\partial T}\right)_{\Omega}
        =-\frac{8\pi^2 l T}{(1-l^2 \Omega^2)}<0,
\end{equation}
which means that $\mu$ is monotonically decreasing with $T$. With the relation
\begin{align*}
        \mu|_{T=0}=0,
\end{align*}
which implies that $\mu$ is identically negative. For the rotating BTZ black holes in the restricted phase space, there is neither an inflection point nor an extremal point in the $T-S$ and $\Omega-J$ sub-phase spaces. So, the phase transition is inexistence in the thermodynamics of the rotating BTZ black holes.

What's more, we consider the heat capacity of the rotating BTZ black hole at constant angular momentum $J$ to study its thermodynamic stability
 \begin{align*}
 C_{J}=T\left(\frac{\partial S}{\partial T}\right)_{J}=\frac{S^{5}-64J^{2}\pi^{4}SC^{2}}{S^{4}+192J^{2}\pi^{4}C^{2}}.
 \end{align*}
 Similarly, we see that the heat capacity $C_{J}$ is also a first order homogeneous function of $S,J,C$, which is expected in extensive thermodynamics. Moreover, the non-negative temperature $T$ constrains $S^{4}\ge64J^{2}\pi^{4}C^{2}$ from Eq.(\ref{E10}), giving a non-negative heat capacity, i.e., $~ C_{J}\ge 0$ always, which indicates that rotating BTZ black holes are thermodynamically stable. In addition, we see that the heat capacity $C_{J}$ depends on $S$ explicitly, while depends on $T$ implicitly via Eq.(\ref{E10}). The intriguing thing is that in the high temperature limit, $C_{J}$ is asymptotically valued as 
 \begin{equation}
        \lim _{t \rightarrow \infty} C_{J} \sim8 \pi^{2} l C T ,
\end{equation}
which is independent of angular momentum $J$. This asymptotic behavior of heat capacity is consistent with the conjecture $C\sim T^{d-2}$ in $d$-dimensional spacetime\cite{Gao:2021xtt}, where $d-2$ is the spatial dimension of the event horizon. It reminds us that Debye’s theory for ordinary non-metallic solid matter in low temperature limit, the heat capacity behaves as $C_V \sim T^D$ in spatial $D$ dimensions. The similar behaviors between them deserve more attention for deep understanding.  
 
\subsection{The charged BTZ black hole}
For the case of charged BTZ black hole with $J=0,\, Q\neq0$, the form of the metric function contains a logarithmic term. This feature makes charged BTZ black holes very different from the charged AdS black holes in four and higher dimensions, so we wonder whether there are extensive thermodynamic properties for the charged BTZ black holes in the restricted formalism. 
From Eq.(\ref{14}), the metric function of charged BTZ black hole is 
 \begin{align*}
 f(r)=-8 G M+\frac{r^{2}}{l^{2}}-4 G Q^{2} \ln \left(\frac{r}{l}\right),~~~
 \Phi(r)=-Q\ln\left(\frac{r}{l}\right).
 \end{align*}
 As before, the thermodynamic variables of the charged BTZ black holes are obtained as
 \begin{align*}
 &M=\frac{r_{0}^{2}}{8 G l^{2}}-\frac{Q^{2}}{2} \ln \left(\frac{r_{0}}{l}\right), \\
 &T=\frac{f'(r_{0})}{4\pi}=\frac{r_{0}}{2 \pi l^{2}}-\frac{G Q^{2}}{\pi r_{0}}, \\
 &S=\frac{\mathcal{A}}{4G}=\frac{\pi r_{0}}{2 G},
 \end{align*}
Similar to the previous subsection, the on-shell action of the charged BTZ black hole is given by\cite{Dhumuntarao:2021xxo}
\begin{align*}
&I_{E H}=-\frac{1}{2 \kappa} \int d^{3} x \sqrt{-g}\left(R-2\Lambda\right)+\frac{1}{4 \mu_{0}} \int d^{3} x F_{\mu\nu} F^{\mu\nu}, \\
&I_{G HY}=-\frac{1}{\kappa} \int_{\partial M} d^{2} x \sqrt{h} K-\frac{1}{\mu_{0}} \int_{\partial M} d^{2} x n_{r} F^{r t} A_{t}, \\
&I_{c t}=\frac{1}{\kappa} \int_{\partial M} d^{2} x \sqrt{h} (\frac{1}{l}).
\end{align*}
Sum up all these contributions, the result is
$$
I_{E}=I_{E H}+I_{G HY}+I_{c t}=\beta \left(\frac{Q^2}{2}\ln(\frac{r_{0}}{l})+\frac{Q^2}{2}-\frac{r_{0}^2}{8Gl^2}\right),
$$
from which, we can prove $
\mu C=TI_{E}=M-TS-\hat{\Phi} \hat{Q}$ with $T=1/\beta$, 
where $\hat{\Phi}=\frac{\Phi \sqrt{G}}{l}, \hat{Q}=\frac{Q l}{\sqrt{G}}$
\footnote{The definition of the charge and potential 
are different with the holographic dictionary $\tilde{\Phi}=\Phi /l,\tilde{Q}=Ql$ \cite{Karch:2015rpa,Chamblin:1999tk}, since bulk action Eq.(\ref{04}) in this paper is slightly different in the coefficients with the usual bulk action $ I=\frac{1}{16\pi G}\int d^{d} x \sqrt{-g}\left(R-2\Lambda-F^{2}\right)$.}. From the expressions of thermodynamic variables, the first law of black hole thermodynamics and the Euler relation are easily verified
 \begin{align}
 \mathrm{d} M &=T \mathrm{d} S+\hat{\Phi} \mathrm{d} \hat{Q}+\mu \mathrm{d} C, \label{33}\\
 M &=T S+\hat{\Phi} \hat{Q}+\mu C \label{34}.
 \end{align}
Here the AdS radius $l$ is still restricted and Newton constant is allowed to vary. 

Also we want to prove that there exist extensive and intensive thermal quantities in this restricted formalism of charged BTZ black hole. According to Eq.(\ref{E20}), the mass $M$ of the charged BTZ black hole is rewritten as
 \begin{equation}
 M=\frac{S^{2}}{16l \pi^{2} C}-\frac{\hat{Q}^{2} \ln \left[S/{(4 \pi C)}\right]}{16 l C}.
 \end{equation}
 And other thermodynamic variables are as follows,
 \begin{align}
 T&=\left(\frac{\partial M}{\partial S}\right)_{\hat{Q},C}=-\frac{\hat{Q}^{2}}{16 l S C}+\frac{S}{8l \pi^{2} C}, \label{45} \\
 \hat{\Phi}&=\left(\frac{\partial M}{\partial \hat{Q}}\right)_{S,C}=-\frac{\hat{Q} \ln \left[S/{(4 \pi C)}\right]}{8l C},  \label{46}\\
 \mu&=\left(\frac{\partial M}{\partial C}\right)_{\hat{Q},S}=\frac{\hat{Q}^{2}}{16l C^{2}}-\frac{S^{2}}{16 \pi^{2}l C^{2}}+\frac{\hat{Q}^{2} \ln \left[S/{(4 \pi C)}\right]}{16l C^{2}}.\label{47}
 \end{align}
When $S \to \lambda S,\hat{Q} \to \lambda \hat{Q},C \to \lambda C$, it is clear that $M$ scales as $M \to \lambda M$, while $T,\hat{\Phi},\mu$ are not rescaled. This proves the first order homogeneity of $M$ and zeroth order homogeneity of  $T,\hat{\Phi},\mu$ in $S,\hat{Q},C$. 
 
 Now we prove that $S$ and $\hat Q$ are extensive variables.
 From Eq.(\ref{46}) and Eq.(\ref{47}), we have
 \begin{equation}
 \mu=\frac{\hat{Q}^{2}}{16l C^{2}}-\frac{S^{2}}{16 l \pi^{2} C^{2}}-\frac{\hat{Q} \hat{\Phi}}{2 C}.\label{E39}
 \end{equation}
 From Eq.(\ref{E39}), we obtain physical expression for $\hat{Q}$
 \begin{equation*}
 \hat{Q}=\frac{4 \pi C \hat{\Phi} l+\sqrt{S^{2}+16 \pi^{2}l C^{2}\left(\mu+l \hat{\Phi}^{2}\right)}}{\pi}.\label{48}
 \end{equation*}
Assuming $S=C\mathcal{S}$, the expression for $\hat{Q}$ reduce to
\begin{align}
 \hat Q=C\mathcal{\hat Q},~~~\mathcal{\hat Q}=4 \hat{\Phi} l+\sqrt{\left[\mathcal{S}^{2}+16 \pi^{2}l\left(\mu+l \hat{\Phi}^{2}\right)\right]\pi}.\label{50}
 \end{align}
 Taking Eq.(\ref{50}) into Eq.(\ref{45}), excepting $S<0$, we obtain the following expression
 \begin{align}
S=C\mathcal{S},~~~\mathcal{S}=4\pi^2lT+\pi \sqrt{(\mathcal{\hat Q}^2+32\pi^2l^2T^2)/2}. \label{43}
 \end{align}
Combining Eq.(\ref{50}) and Eq.(\ref{43}), the expressions of $\mathcal{\hat Q}, \mathcal{S}$ can be solved directly in principle. However, the formulas are very tedious, and the explicit form are unnecessary here, so we do not show these in detail. Indeed, it is obvious that $\mathcal{\hat Q}, \mathcal{S}$ can be determined from the implicit functions, which show us that they only depend on the intensive variables $T$, $\hat{\Phi}$ and $\mu$. Therefore, $S,\hat Q$ are proved to be extensive variables proportionally to $C$.
 
The Gibbs-Duhem equation derived from Eq.(\ref{33}) and Eq.(\ref{34}) is
 \begin{align*}
     \mathrm{d} \mu=-\hat{\mathcal{Q}} \mathrm{d} \hat{\Phi}-\mathcal{S} \mathrm{d} T,
 \end{align*}
 where
 $\hat{\mathcal{Q}}=\hat{Q} / C$ and $ \mathcal{S}=S / C$ are zeroth order homogeneous functions in $S, \hat{Q}, C$. It should be noticed that the chemical potential $\mu$ depend only on $T, ~\hat{\Phi}$ decided by Eqs.(\ref{45}-\ref{47}). We are concerning the behaviors of $\mu-T$ with fixed $\hat{\Phi}$ to investigate the possible phase transition. The directly calculation of the relation 
$\mu(T,\hat{\Phi}$) is difficult, so we derive it with the help of the chain rule of partial derivative
 \begin{align*}
     \left(\frac{\partial \mu}{\partial T}\right)_{\hat{\Phi}}
     &=\left(\frac{\partial \mu}{\partial \hat{Q}}\right)_{S} \left(\frac{\partial S}{\partial T}\right)_{\hat{\Phi}}
     +\left(\frac{\partial \mu}{\partial S}\right)_{\hat{Q}} \left(\frac{\partial \hat{Q}}{\partial T}\right)_{\hat{\Phi}}\\
     &=\frac{4\pi \hat{Q}^3(1+\ln(\frac{S}{4\pi C}))e^{8lC\hat{\Phi}/\hat{Q}}}{128lC^3\hat{\Phi}-\hat{Q}^2(\hat{Q}-4lC\hat{\Phi})e^{16lC\hat{\Phi}/\hat{Q}}}\\
     &+\frac{S \left(\pi ^2 \hat{Q}^2-2 S^2\right) \left(\ln \left(\frac{S}{4 \pi  C}\right)\right)^3}{2 \left(64 \pi ^2 C^3 l^2 \hat{\Phi} ^2+32 \pi ^2 C^3 l^2 \Phi ^2 \ln \left(\frac{S}{4 \pi  C}\right)+C S^2 \left(\ln \left(\frac{S}{4 \pi  C}\right)\right)^3\right)}.
 \end{align*}
Although the above relation seems complicated for us to analysis, while $\hat{\Phi}<0$ from Eq.(\ref{45}) and $2S^{2}\ge\pi^{2}{\hat{Q}^{2}}$ from the requirement of non-negative temperature explicitly show that $\left(\frac{\partial \mu}{\partial T}\right)_{\hat{\Phi}}<0$. It implies that the chemical potential is monotonically decreasing with $T$, and there is no swallow tail behaviors as that in Van der Walls systems. For charged BTZ black holes, we also verify that neither an extremal point nor an inflection point exist in the $T-S$ or $\hat{\Phi}-\hat Q$ phase space, which suggests that there is no second order phase transition.

The heat capacity at constant charge of the charged BTZ black holes is given by
  \begin{align*}
  &C_{\hat Q}=T\left(\frac{\partial S}{\partial T}\right)_{\hat Q}=\frac{S\left(-\pi^{2}{\hat{Q}^{2}}+2S^{2}\right)}{\pi^{2}{\hat{Q}^{2}+2S^{2}}}.
  \end{align*}
The requirement of Hawking temperature $T\ge0$, i.e., $2S^{2}\ge\pi^{2}{\hat{Q}^{2}}$, 
implies that the heat capacity $C_{\hat Q}\ge 0$ and the charged BTZ black hole is always thermodynamically stable. Besides, the heat capacity $C_{\hat Q}$ is also a first order homogeneous function of $S,\hat{Q},C$ as we expected. In high temperature asymptotic region, the value of $C_{\hat Q}$ is given by
\begin{equation}
        \lim _{t \rightarrow \infty} C_{\hat{Q}} \sim8 \pi^{2} l C T,
\end{equation}
which is independent of the charge $Q$, and is the same as that of rotating BTZ black hole.
 
  \section{Conclusion}\label{III}
In this paper, the thermodynamic properties of the three dimensional BTZ black holes are restudied motivated by the recent viewpoint on holographic thermodynamics. In the previous extended phase space formalism, we notice that there are some puzzling problems for the Smarr relation of the three dimensional BTZ black holes: $1)$ The mass term and the terms related to the electric field disappear; $2)$ The thermodynamic potential does not satisfy the Euler homogeneity, which is crucial for understanding the equilibrium state of thermodynamics. 

To address the above problems, we attempt to restrict the AdS radius $l$ as a constant and vary the Newtonian constant $G$. In this formalism, a pair of new conjugate variables $(\mu, C)$ is introduced brought from the AdS/CFT correspondence. The resulting thermodynamics are compatible with standard extensive thermodynamics, all the thermodynamic quantities are classified into extensive and intensive ones. The entropy $S$ and angular momentum $J$ (or charge $\hat{Q}$) are proved to be extensive, and thus they are simple additive. The other thermal quantities $T, \, \Omega, \,\mu$ are intensive. The black hole mass $M$ is regarded as internal energy and is a first order homogeneous function of the relevant extensive variables. We find the first law and Euler relation as well as the Gibbs-Duhem equation of the BTZ black holes hold simultaneously.

In addition to this, the phase spaces $T-S$ and $\Omega-J$ of the rotating BTZ black holes (or the phase spaces $T-S$ and $\hat \Phi-\hat Q$ of the charged BTZ black holes), constructed by the independent variables, are investigated and we find that there is neither an extremum nor an inflection point. Thus, the rotating and charged BTZ black holes have no critical phenomena in this formalism, which is similar to the phase structure of BTZ black holes in extended phase space. Of course, we also find that both rotating and charged BTZ black holes are thermodynamically
stable, which is determined by the non-negativity of the corresponding heat capacity.

There are many coupling parameters in modified gravity theory models, and usually researchers have different opinions on whether these parameters can be treated as thermodynamic variables. For the requirements on the extension of thermodynamics, what constraints will be imposed on these coupling parameters is an interesting issue that we want to investigate in the near future.

\section{Acknowledgement} 
Bin Wu would like to thank Prof. Liu Zhao for the useful discussion. This work is supported by the National Natural Science Foundation of China (Grant Nos.12047502, 11947208), the China Postdoctoral Science Foundation (Grant Nos.2017M623219). 
\end{spacing}

\end{document}